\documentclass[onecollarge]{svjour2}
\usepackage{amsmath,amssymb,bm,graphicx,hyperref,multirow,slashbox}
\journalname{Few-Body Systems}

\renewcommand\d{\partial}
\newcommand\<{{\langle}}
\renewcommand\>{{\rangle}}
\renewcommand\r{{\bm{r}}}
\newcommand\x{{\bm{x}}}
\renewcommand\k{{\bm{k}}}
\newcommand\p{{\bm{p}}}
\newcommand\q{{\bm{q}}}
\newcommand\0{{\bm{0}}}
\newcommand\eff{{\mathrm{eff}}}

\begin{document}

\title{Liberating Efimov physics from three dimensions
\thanks{Special issue devoted to Efimov physics}}

\author{Yusuke Nishida \and Shina Tan}

\institute{Yusuke Nishida \at
Center for Theoretical Physics, Massachusetts Institute of Technology,
Cambridge, MA 02139, USA \\
\email{nishida@mit.edu}
\and Shina Tan \at
School of Physics, Georgia Institute of Technology,
Atlanta, GA 30332, USA \\
\email{shina.tan@physics.gatech.edu}}

\date{April 2011}

\maketitle
%\vspace{-87.4mm}{\small\hspace{127mm}MIT-CTP 4245}\vspace{82.8mm}

\begin{abstract}
 When two particles attract via a resonant short-range interaction,
 three particles always form an infinite tower of bound states
 characterized by a discrete scaling symmetry.  It has been considered
 that this Efimov effect exists only in three dimensions.  Here we
 review how the Efimov physics can be liberated from three dimensions by
 considering two-body and three-body interactions in mixed dimensions
 and four-body interaction in one dimension.  In such new systems,
 intriguing phenomena appear, such as confinement-induced Efimov effect,
 Bose-Fermi crossover in Efimov spectrum, and formation of interlayer
 Efimov trimers.  Some of them are observable in ultracold atom
 experiments and we believe that this study significantly broadens our
 horizons of universal Efimov physics.
\end{abstract}

\setcounter{tocdepth}{2}
\renewcommand{\baselinestretch}{1.3}\selectfont
\tableofcontents
\renewcommand{\baselinestretch}{1}\selectfont

\section{Introduction \label{sec:introduction}}
When particles attract via a short-range two-body interaction whose
$s$-wave scattering length is much larger than the range of the
interaction potential, their low-energy physics becomes
{\em universal\/}, i.e., independent of details of the interaction
potential~\cite{Braaten:2004rn}.  In 1970, Vitaly Efimov discovered that
three particles at infinite scattering length form an infinite tower of
bound states whose binding energies $E_n$ have an accumulation point at
$E=0$~\cite{Efimov:1970}.  The most important characteristic of this
Efimov effect is the existence of a discrete scaling symmetry.  Namely,
the ratio of two successive binding energies is constant
\begin{equation}\label{eq:spectrum}
 \frac{E_{n+1}}{E_{n}} = \lambda^{-2}
\end{equation}
where $\lambda$ is a scaling factor and given by $\lambda=22.7$ for
three identical bosons.  This discrete scaling symmetry is descended
from the full scaling symmetry of the zero-range two-body interaction at
infinite scattering length.  The necessity to regularize the
short-distance behavior of three particles breaks the full scaling
symmetry down to a discrete one and leads to the infinite tower of bound
states.  This peculiar phenomenon is a rare manifestation of the
renormalization group limit cycle in physics~\cite{Bedaque:1998kg}.  In
this article, {\em we define the Efimov effect as the appearance of an
infinite tower of bound states with a discrete scaling symmetry when
particles attract via short-range interactions.\/}

After the discovery by Efimov, it has been repeatedly discussed that the
Efimov effect is absent in two
dimensions~\cite{Bruch:1979,Lim:1980a,Lim:1980b,Vugalter:1983,Adhikari:1986,Adhikari:1988,Nielsen:1997}.
More precisely, the Efimov effect for three identical bosons appears
only in an interval of $2.30<d<3.76$~\cite{Nielsen:2001}.  Why does the
Efimov effect exist only in three dimensions?  Or, can we liberate the
universal Efimov physics from three dimensions?  More generally, when
does such a peculiar phenomenon appear?  These are questions we would
like to address in this article.

\begin{table}[b]
 \renewcommand{\arraystretch}{1.4}
 \caption{Efimov effect in $AAB$ systems with resonant two-body
 interactions between $A$ and $B$ particles in pure 3D, 2D-3D, 1D-3D,
 2D-2D, and 1D-2D mixtures.  $\circ$ ($\times$) indicates the presence
 (absence) of the Efimov effect for any mass ratio $m_A/m_B$, while
 numbers indicate critical mass ratios above which the Efimov effect
 appears.  \label{tab:2-body}}
 \centering
 \begin{tabular}{c|cc|cc|cc}
  \multirow{2}{*}{\backslashbox{$B$}{\\$A$}}
  & \multicolumn{2}{c}{3D$\ \ $} & \multicolumn{2}{|c|}{2D$\ \ $}
  & \multicolumn{2}{c}{1D$\ \ $} \\
  & boson & fermion & boson & fermion & boson & fermion \\\hline
  3D & $\circ$ & $13.6$ & $\circ$ & $6.35$ & $\circ$ & $2.06$ \\
  2D & $\circ$ & $28.5$ & $\circ$ & $11.0$ & $\circ$ & $\times$ \\
  1D & $\circ$ & $155$ & $\circ$ & $\times$ 
  & \multicolumn{2}{c}{---$\ $}
 \end{tabular}
\end{table}

In Sect.~\ref{sec:absence}, we provide a simple argument to understand
the absence of the Efimov effect in other than three dimensions.  Our
logic is such that {\em in order for the Efimov effect to appear, it is
necessary that particles attract via scale-invariant interactions\/}
from the definition of the Efimov effect.  However, short-range two-body
interactions can satisfy this necessary condition only in three
dimensions.  Therefore, the key to liberate the Efimov physics from
three dimensions is whether we can find scale-invariant short-range
interactions in other than three dimensions.  In
Sect.~\ref{sec:unitarity}, we show that two-body and three-body
interactions in mixed dimensions and four-body interaction in one
dimension can be scale invariant at their resonances and thus they
satisfy the necessary condition for the Efimov effect.  All such new
systems indeed exhibit the Efimov effect as we will show in a separate
paper~\cite{Tan-Nishida}.  Table~\ref{tab:2-body} summaries a part of
our results on the presence of the Efimov effect and critical mass
ratios in various three-body systems with resonant two-body
interactions.  In Sect.~\ref{sec:efimov}, we discuss a number of
universal phenomena unique in such new systems and observable in
ultracold atom experiments.  Section~\ref{sec:summary} is the summary of
this article and some details of results discussed in the text are
presented in Appendices.  In view of the fact that most studies on the
Efimov effect have been devoted to three dimensions for more than 40
years, we believe that this study significantly broadens our horizons of
universal Efimov physics.

\section{Absence of Efimov effect in other than three dimensions
 \label{sec:absence}}
Suppose two particles interact via a short-range potential $V(\r)$ that
vanishes outside its potential range; $V(|\r|>r_0)=0$.  After separating
the center-of-mass motions, the two-body scattering is described by the
three-dimensional Schr\"odinger equation
\begin{equation}\label{eq:schrodinger}
 \left[-\frac{\hbar^2}{2\mu}\frac{\d^2}{\d\r^2} + V(\r)\right]\psi(\r)
 = E\,\psi(\r) \qquad \text{with} \quad \r=(r_1,r_2,r_3),
\end{equation}
where $\mu$ is the reduced mass.  The relative wave function outside the
potential range but at the distance scale much shorter than the de
Broglie wavelength, $r_0<|\r|\ll\hbar/\sqrt{2\mu E}$, obeys the
Laplace's equation
\begin{equation}\label{eq:laplace}
 \frac{\d^2}{\d\r^2}\psi(\r) = 0.
\end{equation}
The general solution to this equation in the $s$-wave channel is given
by
\begin{equation}
 \psi(\r) \propto \frac1{|\r|} - \frac1{a},
\end{equation}
where $a$ is the $s$-wave scattering length.  The scattering length
becomes infinite $a\to\infty$ at the resonance where the interaction
potential supports a two-body bound state exactly at zero binding
energy.  Furthermore, if we are interested in the low-energy physics
whose de Broglie wavelength is much longer than the potential range
$r_0\ll\hbar/\sqrt{2\mu E}$, we can regard such a short-range potential
as a zero-range potential.  The scale invariance is achieved by taking
the zero-range limit $r_0\to0$ with keeping the scattering length
infinite.  Now the Schr\"odinger equation becomes
\begin{equation}
 -\frac{\hbar^2}{2\mu}\frac{\d^2}{\d\r^2}\psi(\r)
 = E\,\psi(\r) \qquad \text{for} \quad |\r|\neq0
\end{equation}
and the interaction potential is replaced by the short-range boundary
condition:
\begin{equation}\label{eq:short-range}
 \psi(|\r|\to0) \propto \frac1{|\r|} + O(|\r|).
\end{equation}
Then by adding a third particle, one can show that the Efimov effect
appears~\cite{Efimov:1970}.  Our purpose here is not to show the
presence of the Efimov effect in three dimensions but to show the
presence of the scale-invariant interaction, which is necessary for the
Efimov effect.  We refer to the zero-range and infinite scattering
length interaction implemented by Eq.~(\ref{eq:short-range}) (and its
analytic continuation to other spatial dimensions) as the unitarity
interaction.

In order to understand the absence of the Efimov effect in other than
three dimensions, we generalize the above argument to arbitrary spatial
dimensions.  In $D$ dimensions, the short-range boundary condition
(\ref{eq:short-range}) is modified as
\begin{equation}\label{eq:general_d}
 \psi(|\r|\to0) \propto \frac1{|\r|^{D-2}} + O(|\r|^{4-D},|\r|^2),
\end{equation}
which is the singular solution to the $D$-dimensional Laplace's equation
(\ref{eq:laplace}).  Because the wave function has to be normalizable,
the singularity of the wave function in Eq.~(\ref{eq:general_d}) is
unacceptable in higher dimensions $D>4$.  In $D=4$, the normalization
integral is logarithmically divergent at the origin which implies that
two particles behave as a point-like composite
particle~\cite{Nussinov:2006zz,Nishida:2010tm}.  This composite particle
is noninteracting with a third particle and thus they cannot form bound
states.  On the other hand, the singularity of the wave function
disappears in $D=2$ which means that the interaction between two
particles also disappears.  Obviously, three particles cannot form bound
states from no interaction.  Finally in $D=1$, the wave function
(\ref{eq:general_d}) behaves as $\psi(|\r|)\propto|\r|$ at the short
distance, which is a consequence of the hardcore repulsion between two
particles.  The hardcore repulsion is scale invariant but three
particles cannot form bound states from the repulsive interaction.
Indeed, the spectrum of particles interacting via the hardcore repulsion
in $D=1$ is equivalent to that of noninteracting identical
fermions~\cite{Girardeau:1960}.  Therefore, we find that the Efimov
effect cannot appear in one, two, and four dimensions from any
short-range two-body interactions because the unitarity interaction
becomes trivial in such dimensions.  Table~\ref{tab:three_bosons}
summarizes the fate of three identical bosons in the unitarity limit as
a function of the dimensionality $D$.

\begin{table}[t]
 \renewcommand{\arraystretch}{1.8}
 \caption{Fate of three identical bosons in the unitarity limit.  They
 form an Efimov trimer only in $D=3$ and become trivial systems in
 $D=1,2,4$.  The unitarity interaction does not exist in higher
 dimensions $D>4$. \label{tab:three_bosons}}
 \centering\smallskip
 \begin{tabular}{ll}\hline
  \,$D=1$ & hardcore repulsion \,$\Rightarrow$\, free fermions \\\hline
  \,$D=2$ & free bosons \\\hline
  $\bm{D=3}$ & {\bf Efimov trimer} \ ($2.30<d<3.76$) \\\hline
  \,$D=4$ & free composite $+$ one boson \\\hline
 \end{tabular}
\end{table}

For completeness, we also consider higher partial-wave interactions.  In
a partial-wave channel with orbital angular momentum $\ell\geq0$, the
radial wave function at the distance scale
$r_0<|\r|\ll\hbar/\sqrt{2\mu E}$ behaves as
\begin{equation}
 \psi(\r) \propto \frac1{|\r|^{\ell+D-2}} - \frac{|\r|^\ell}{a_\ell}.
\end{equation}
In order for the interaction potential to produce the scale-invariant
attraction in the zero-range limit $r_0\to0$, the power of the first
term needs to satisfy
\begin{equation}\label{eq:constraint}
 0 < \ell+D-2 < \frac{D}2.
\end{equation}
The first inequality comes from the requirement of the attraction and
the second comes from the normalizability of the wave function.  The
constraint (\ref{eq:constraint}) cannot be satisfied by any non-negative
integers $\ell=0,1,2,\dots$ in $D\geq4$, while it can be satisfied only
in the $s$-wave channel $\ell=0$ in $D=3$ that is the case discussed
above.  In $D=2$, Eq.~(\ref{eq:constraint}) reduces to $0<\ell<1$, which
cannot be satisfied by bosons ($\ell=0$) or fermions ($\ell=1$) but can
be satisfied by anyons~\cite{Nishida:2007de}.  The case of anyons will
be treated separately in Sect.~\ref{sec:anyon}.

\section[Unitarity interactions beyond three dimensions]
 {Unitarity interactions beyond three dimensions \cite{Nishida:2008kr}
 \label{sec:unitarity}}
We have seen that the Efimov effect can appear only in three dimensions
because the scale-invariant unitarity interaction is nontrivial only in
three dimensions and becomes trivial in one, two, and four dimensions
and does not exist in higher dimensions.  Therefore, the key to liberate
the Efimov physics from three dimensions is whether we can find
nontrivial scale-invariant short-range interactions in other than three
dimensions.  We recall that three dimensions were special because the
relative wave function at a short distance obeys the three-dimensional
Laplace's equation (\ref{eq:laplace}) that admits the normalizable
singular solution as in Eq.~(\ref{eq:short-range}).  However, three
spatial dimensions are actually not essential for the three-dimensional
Laplace's equation.  What is essential is the fact that the scattering
is described by three relative coordinates $\r=(r_1,r_2,r_3)$ and their
physical meaning does not matter.  Of course, as far as the scattering
of two particles living in the same space is concerned, the numbers of
spatial dimensions and relative coordinates coincide, but more generally
they do not.  For example, the scattering of four particles in one
dimension is described by the same Schr\"odinger equation as
Eq.~(\ref{eq:schrodinger}) with three relative coordinates
$\r=(r_1,r_2,r_3)$ because each particle has one coordinate and one
center-of-mass coordinate can be separated.  The short-range potential
$V(\r)$ in this case represents a four-body interaction in one
dimension, which produces the scale-invariant attraction in the
zero-range limit at the four-body resonance.

In order to elucidate all possible scale-invariant short-range
interactions, we consider the scattering of $N$ particles labeled by
$i=1,\dots,N$, each of which lives in a $d_i$-dimensional flat space.
If all spaces have a common $d_\parallel$-dimensional intersection, the
center-of-mass motions in such directions can be separated from the
relative motions.  Therefore, by introducing the codimension
$d_{\perp i}\equiv d_i-d_\parallel$, the number of relative coordinates
is given by
\begin{equation}
 D = \sum_{i=1}^Nd_i - d_\parallel
  = \sum_{i=1}^Nd_{\perp i} + \left(N-1\right)d_\parallel
\end{equation}
and the scattering problem is described by the $D$-dimensional
Schr\"odinger equation (\ref{eq:schrodinger}).  If $d_{\perp i}=0$ for
all $i$, all particles live in the same space while otherwise they live
in different spaces and interact only at their intersection.  The latter
case is referred to as {\em mixed dimensions}.  As we have shown, the
short-range $N$-body interaction potential $V(\r)$ can produce the
scale-invariant attraction in the unitarity limit only when $D=3$.  All
possible combinations of spatial dimensions for $D=3$ are summarized in
Table~\ref{tab:mixed_dimensions}.  Here we assumed $d_i\geq1$ for all
$i$ because $d_i=0$ corresponds a zero-dimensional impurity and is
rather trivial.  In addition to the ordinary two-body interaction in
three dimensions, we find seven more systems in which short-range
few-body interactions become scale invariant in the unitarity limit;
two-body and three-body interactions in mixed dimensions and four-body
interaction in one dimension~\cite{Nishida:2008kr}.  These new systems
in the unitarity limit satisfy the necessary condition for the Efimov
effect and thus we expect that the addition of an $(N{+}1)$th particle
may lead to the formation of an infinite tower of $(N{+}1)$-body bound
states with a discrete scaling symmetry.  Indeed, this expectation turns
out to be true and {\em$N{+}1$ particles in all such new systems exhibit
the Efimov effect\/} as we will show in a separate
paper~\cite{Tan-Nishida}.  By this way, we can liberate the universal
Efimov physics from three dimensions\,!

\begin{table}[t]
 \renewcommand{\arraystretch}{1.6}
 \caption{All possible combinations of spatial dimensions $d_i\geq1$
 where the scattering of $N$ particles is described by three relative
 coordinates~\cite{Nishida:2008kr}.  In each system, an $N$-body
 interaction potential becomes scale invariant in the unitarity limit
 and $N{+}1$ particles exhibit the Efimov effect.
 \label{tab:mixed_dimensions}}
 \centering\smallskip
 \begin{tabular}{lll}\hline
  \# of particles and dimensions & Name & Example of geometry \\\hline
  $N=2$ \ \ $(d_A,d_B;d_\parallel)=(3,3;3)$
  & pure 3D & $\x_A=\x_B=(x,y,z)$ \\
  $N=2$ \ \ $(d_A,d_B;d_\parallel)=(2,3;2)$
  & 2D-3D mixture & $\x_A=(x,y)$ \quad $\x_B=(x,y,z)$ \\
  $N=2$ \ \ $(d_A,d_B;d_\parallel)=(1,3;1)$
  & 1D-3D mixture & $\x_A=(z)$ \quad $\x_B=(x,y,z)$ \\
  $N=2$ \ \ $(d_A,d_B;d_\parallel)=(2,2;1)$
  & 2D-2D mixture & $\x_A=(x,z)$ \quad $\x_B=(y,z)$ \\
  $N=2$ \ \ $(d_A,d_B;d_\parallel)=(1,2;0)$
  & 1D-2D mixture & $\x_A=(z)$ \quad $\x_B=(x,y)$ \\
  $N=3$ \ \ $(d_A,d_B,d_C;d_\parallel)=(1,1,2;1)$
  & 1D$^2$-2D mixture & $\x_A=\x_B=(x)$ \quad $\x_C=(x,y)$ \\
  $N=3$ \ \ $(d_A,d_B,d_C;d_\parallel)=(1,1,1;0)$
  & 1D-1D-1D mixture \ & $\x_A=(x)$ \quad $\x_B=(y)$ \quad $\x_C=(z)$ \\
  $N=4$ \ \ $(d_A,d_B,d_C,d_D;d_\parallel)=(1,1,1,1;1)$ \ 
  & pure 1D & $\x_A=\x_B=\x_C=\x_D=(x)$ \\\hline
 \end{tabular}
\end{table}

\section{Efimov physics beyond three dimensions \label{sec:efimov}}
We shall defer to Ref.~\cite{Tan-Nishida} on complete and unified
analyses of the Efimov effect in all possible cases found in the
previous section.  Instead, here we focus on a number of universal
phenomena that cannot be found in the ordinary Efimov physics in three
dimensions and are thus unique in our new systems.  Some of them are
observable in ultracold atom experiments.

\subsection[Confinement-induced Efimov effect]
  {Confinement-induced Efimov effect \cite{Nishida:2009fs}
\label{sec:confinement}}
\begin{figure*}
 \includegraphics[width=0.48\textwidth]{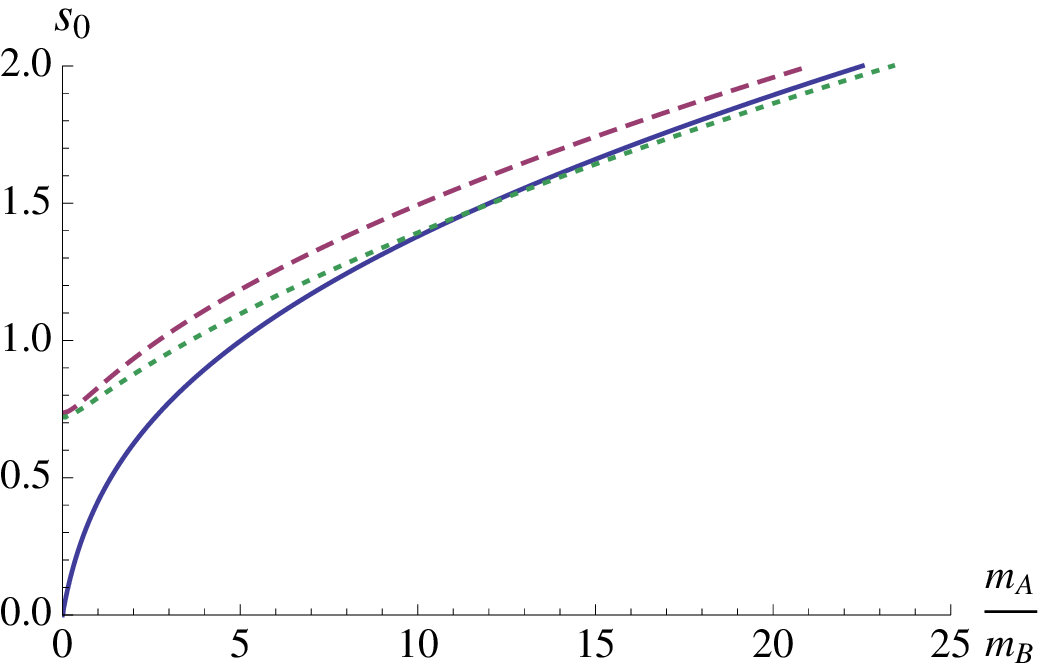}\hfill
 \includegraphics[width=0.48\textwidth]{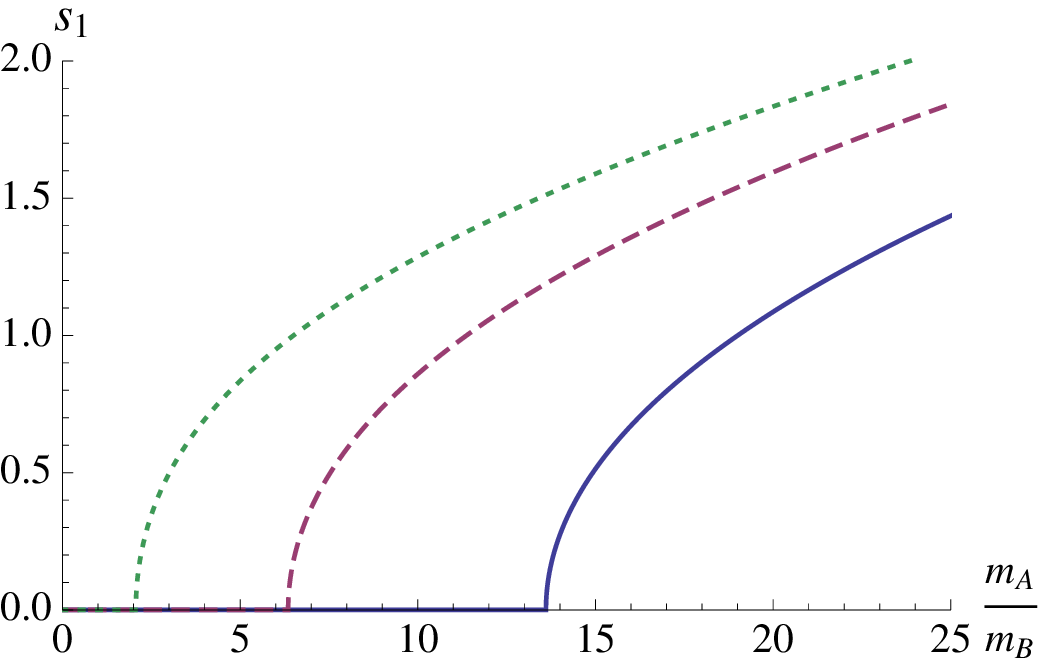}\smallskip
 \caption{Exponents $s_\ell$ of $AAB$ Efimov trimers as functions of the
 mass ratio $m_A/m_B$ when $A$ particles are bosons (left) and fermions
 (right).  Solid, dashed, and dotted curves correspond to pure 3D, 2D-3D
 mixture, and 1D-3D mixture, respectively, in which two $A$ particles
 are confined in lower dimensions and resonantly interact with one $B$
 particle in 3D.  Scaling factors are given by $\lambda=e^{\pi/s_\ell}$
 and $s_\ell=0$ indicates the absence of the Efimov effect.
 \label{fig:scaling_AAB}}
\end{figure*}

Let us consider a three-body system composed of two $A$ and one $B$
particles in which $A$ and $B$ particles attract via a zero-range and
infinite scattering length interaction while two $A$ particles do not
interact with each other.  The presence or absence of the Efimov effect
can be diagnosed by studying the scaling behavior of three-body wave
function $\psi\sim R^{\gamma_\ell}$ at a small hyperradius $R\to0$ in a
given partial-wave channel classified by a quantum number
$\ell$.\footnote{For example, $\ell=0,1,2,\dots$ is the orbital angular
momentum in pure 3D, $\ell=0,\pm1,\pm2,\dots$ is the magnetic quantum
number in 2D-3D mixture, and $(-1)^\ell=\pm1$\,\ ($\ell=0,1$) is the
parity in 1D-3D mixture.}  When the scaling exponent $\gamma_\ell$
develops an imaginary part $\mathrm{Im}[\gamma_\ell]\equiv s_\ell>0$,
the Efimov effect appears in the $\ell$th partial-wave
channel\footnote{On the other hand, real solutions of the scaling
exponent $\gamma_\ell$ correspond to energies of $N{+}1$ particles in a
harmonic potential by
$E=\left(\frac12\sum_{i=1}^{N+1}d_i+\gamma_\ell+2n\right)\hbar\omega$
with $n=0,1,2,\dots$~\cite{Tan:2004,Werner:2006zz,Nishida:2007pj}.}
and the spectrum of $AAB$ Efimov trimers exhibits the discrete scaling
symmetry as in Eq.~(\ref{eq:spectrum}):
\begin{equation}
 \frac{E_{n+1}}{E_n} = e^{-2\pi/s_\ell} \equiv \lambda^{-2}.
\end{equation}
Here the scaling factor $\lambda=e^{\pi/s_\ell}>1$ depends on the mass
ratio $m_A/m_B$, statistics, and dimensionality of $A$ and $B$
particles.  Figure~\ref{fig:scaling_AAB} shows the exponent $s_\ell$ as
a function of $m_A/m_B$ in pure 3D, 2D-3D mixture, and 1D-3D mixture, in
which two $A$ particles are confined in lower dimensions with keeping
one $B$ particle in 3D~\cite{Nishida:2008kr}.  When $s_\ell$ is larger
(smaller), the spectrum becomes denser (sparser) and $s_\ell=0$
indicates the absence of the Efimov effect.

When $A$ particles are bosons (left panel in
Fig.~\ref{fig:scaling_AAB}), the $AAB$ system always exhibits the Efimov
effect in the $s$-wave or even-parity channel ($\ell=0$) for any mass
ratio, but the scaling factor is strongly affected by the dimensionality
of $A$ particles.  For example, for $m_A/m_B=87/41$ corresponding to the
Bose-Bose mixture of $A={}^{87}$Rb and
$B={}^{41}$K~\cite{Barontini:2009}, the scaling factor is given by
\begin{subequations}
 \begin{alignat}{2}
  & \lambda = 131 &\qquad &\text{in pure 3D}, \\[3pt]
  & \lambda = 27.7 &\qquad &\text{in 2D-3D mixture}, \\[3pt]
  & \lambda = 34.6 &\qquad &\text{in 1D-3D mixture},
 \end{alignat}
\end{subequations}
while for $m_A/m_B=41/87$ corresponding to $A={}^{41}$K and
$B={}^{87}$Rb, we find
\begin{subequations}
 \begin{alignat}{2}
  & \lambda = 3.48\times10^5 &\qquad &\text{in pure 3D}, \\[3pt]
  & \lambda = 59.3 &\qquad &\text{in 2D-3D mixture}, \\[3pt]
  & \lambda = 67.6 &\qquad &\text{in 1D-3D mixture}.
 \end{alignat}
\end{subequations}
Therefore, the scaling factor can be significantly decreased by
confining two $A$ particles in lower dimensions, which makes the
spectrum of $AAB$ Efimov trimers much denser.

On the other hand, when $A$ particles are fermions (right panel in
Fig.~\ref{fig:scaling_AAB}), the effect of the confinement is more
striking.  As a consequence of the Fermi statistics, there is a critical
mass ratio above which the $AAB$ system exhibits the Efimov effect in
the $p$-wave or odd-parity channel ($\ell=1$)~\cite{Petrov:2003}.  The
critical mass ratio depends on the dimensionality of $A$ particles and
is given by
\begin{subequations}\label{eq:critical}
 \begin{alignat}{2}
  & m_A/m_B = 13.6 &\qquad &\text{in pure 3D}, \\[3pt]
  & m_A/m_B = 6.35 &\qquad &\text{in 2D-3D mixture}, \\[3pt]
  & m_A/m_B = 2.06 &\qquad &\text{in 1D-3D mixture}.
 \end{alignat}
\end{subequations}
Therefore, although the Fermi-Fermi mixture of $A={}^{40}$K and
$B={}^6$Li with $m_A/m_B=40/6$ does not exhibit the Efimov effect in a
free space, the Efimov effect can be induced by confining two ${}^{40}$K
atoms in lower dimensions because their mass ratio lies in
$6.35<m_A/m_B<13.6$.  The scaling factor in this case is given by
\begin{subequations} 
 \begin{alignat}{2}
  & \text{no Efimov effect} &\qquad &\text{in pure 3D}, \\[3pt]
  & \lambda = 1.78\times10^5 &\qquad &\text{in 2D-3D mixture}, \\[3pt]
  & \lambda = 22.0 &\qquad &\text{in 1D-3D mixture}.
 \end{alignat}
\end{subequations}
This is a new type of the Efimov effect referred to as
{\em confinement-induced Efimov effect\/} and will be observable in
ultracold atom
experiments~\cite{Taglieber:2008,Wille:2008,Voigt:2009,Tiecke:2010,Naik:2010,Trenkwalder:2010,Wu:2011}
through the measurement of the three-body recombination
rate~\cite{Nishida:2009fs}.  As is discussed in
Refs.~\cite{Nishida:2009fs,Levinsen:2009mn}, resulting Efimov trimers
realized in ultracold atoms have a small decay width and thus they are
expected to be long-lived.  On the other hand, for $m_A/m_B=6/40<2.06$
corresponding to $A={}^6$Li and $B={}^{40}$K, the Efimov effect does not
take place in any dimensions.

\begin{figure*}
 \includegraphics[width=0.48\textwidth]{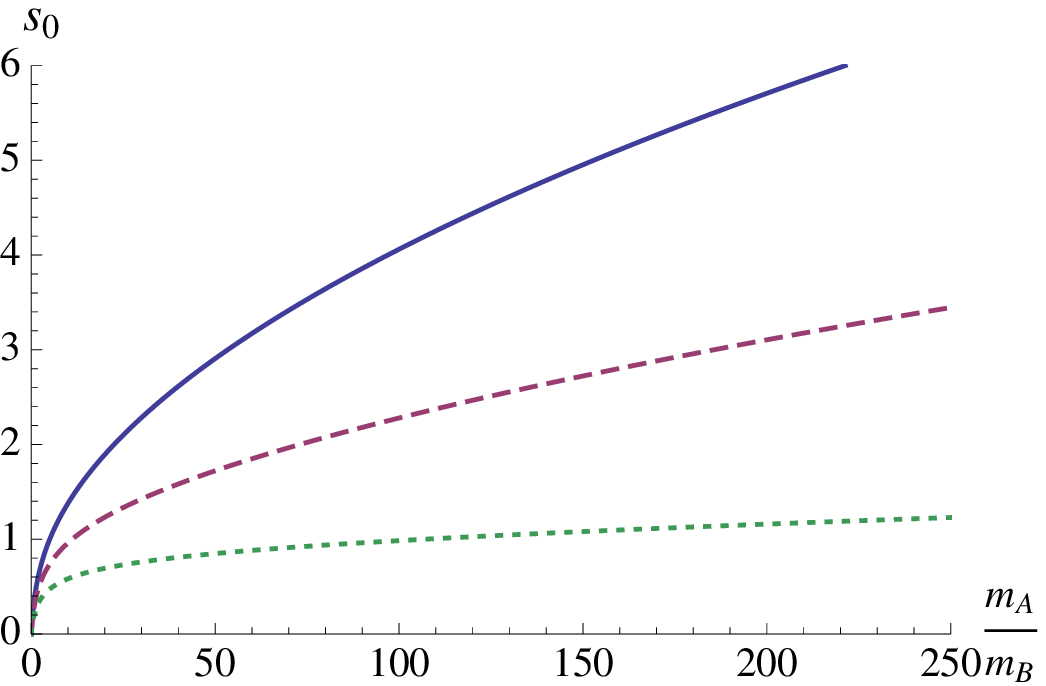}\hfill
 \includegraphics[width=0.48\textwidth]{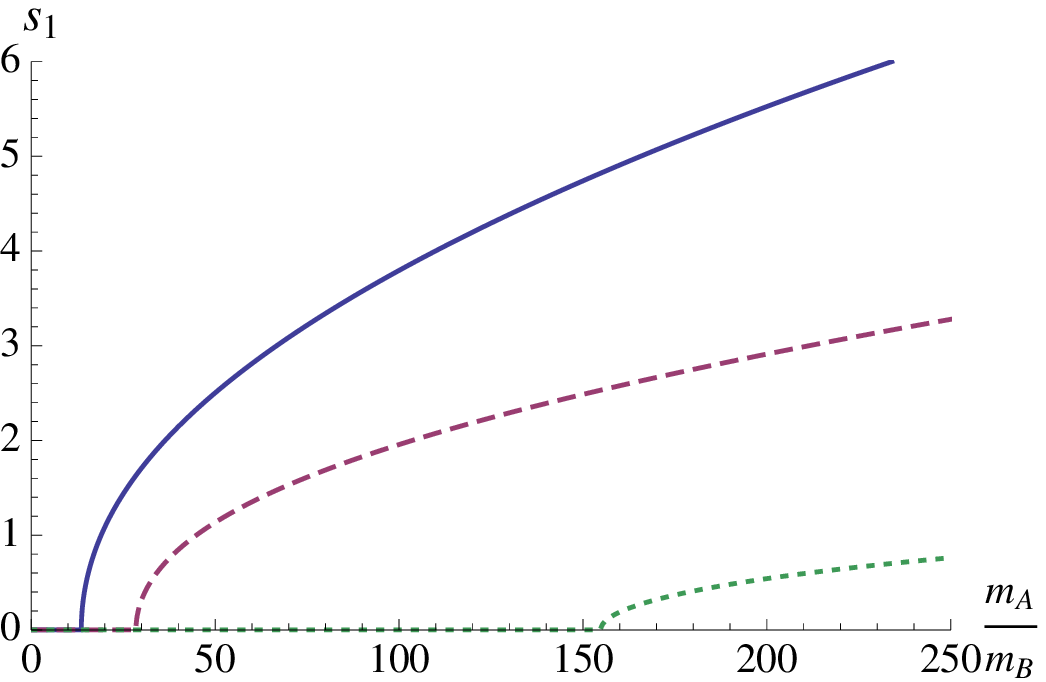}\smallskip
 \caption{Same as Fig.~\ref{fig:scaling_AAB} but for cases in which one
 $B$ particle is confined in lower dimensions and resonantly interacts
 with two $A$ particles in 3D. \label{fig:scaling_ABB}}
\end{figure*}

Figure~\ref{fig:scaling_ABB} shows the same exponent $s_\ell$ but for
cases in which one $B$ particle is confined in lower dimensions with
keeping two $A$ particles in 3D~\cite{Nishida:2008kr}.  When $A$
particles are bosons (left panel in Fig.~\ref{fig:scaling_ABB}), the
$AAB$ system always exhibits the Efimov effect in the $s$-wave or
even-parity channel ($\ell=0$) for any mass ratio $m_A/m_B$, but the
scaling factor $\lambda=e^{\pi/s_\ell}$ is significantly increased by
the confinement.  For example, for $m_A/m_B=87/41$ or $m_A/m_B=41/87$,
we find
\begin{subequations}
 \begin{alignat}{3}
  & \lambda = 131 &\qquad \text{or} \qquad
  & \lambda = 3.48\times10^5 &\qquad &\text{in pure 3D}, \\[3pt]
  & \lambda = 529 &\qquad \text{or} \qquad
  & \lambda = 6.95\times10^6 &\qquad &\text{in 2D-3D mixture}, \\[3pt]
  & \lambda = 1.10\times10^4 &\qquad \text{or} \qquad
  & \lambda = 5.67\times10^9 &\qquad &\text{in 1D-3D mixture},
 \end{alignat}
\end{subequations}
respectively.  On the other hand, when $A$ particles are fermions (right
panel in Fig.~\ref{fig:scaling_ABB}), the $AAB$ system exhibits the
Efimov effect in the $p$-wave or odd-parity channel ($\ell=1$) only
above a critical mass ratio given by
\begin{subequations}
  \begin{alignat}{2}
   & m_A/m_B = 13.6 &\qquad &\text{in pure 3D}, \\[3pt]
   & m_A/m_B = 28.5 &\qquad &\text{in 2D-3D mixture}, \\[3pt]
   & m_A/m_B = 155 &\qquad &\text{in 1D-3D mixture}.
  \end{alignat}
\end{subequations}
Therefore, the formation of $AAB$ Efimov trimers becomes more difficult
by confining one $B$ particle in lower dimensions as opposed to
confining two $A$ particles.  In particular, this Efimov effect does not
take place in any dimensions for the Fermi-Fermi mixture of
$A={}^{40}$K, $B={}^6$Li or $A={}^6$Li, $B={}^{40}$K.

\subsection{Bose-Fermi crossover in Efimov spectrum
  \label{sec:crossover}}
In the above discussions, we assumed that two $A$ particles do not
interact with each other.  The presence of a short-range interaction
between $A$ particles modifies the spectrum of $AAB$ Efimov trimers.
Suppose $A$ particles are bosons and interact with each other via a
zero-range potential with finite scattering length $a$, in addition to
the infinite scattering length interaction with one $B$ particle.
Because $a$ introduces a scale, the spectrum of $AAB$ trimers $\{E_n\}$
is no longer given by Eq.~(\ref{eq:spectrum}).  For deep trimers whose
size $\sim\kappa_n^{-1}\equiv\hbar/\sqrt{2\mu|E_n|}$ is much smaller
than $|a|$, the scattering length can be regarded as effectively
infinite $a\to\infty$.  Because the scale $a$ disappears in such a
limit, the spectrum of trimers exhibits the approximate discrete scaling
symmetry with a certain scaling factor $\lambda_\mathrm{uv}$ due to the
Efimov effect.  On the other hand, for shallow trimers whose size
$\sim\kappa_n^{-1}$ is much larger than $|a|$, the scattering length can
be regarded as effectively vanishing $a\to0$.  Because the scale $a$
disappears in such a limit again, the spectrum of trimers exhibits the
approximate discrete scaling symmetry but with a different scaling
factor $\lambda_\mathrm{ir}$.  Therefore, the spectrum of $AAB$ trimers
shows an intriguing crossover from the ultraviolet scaling behavior
\begin{align}\label{eq:ultraviolet}
 \frac{E_{n+1}}{E_n} &\approx \lambda_\mathrm{uv}^{-2}
 \qquad \text{for} \quad \frac1{\kappa_n|a|}\ll1
 \intertext{to the infrared scaling behavior}
 \frac{E_{n+1}}{E_n} &\approx \lambda_\mathrm{ir}^{-2}
 \qquad \text{for} \quad \frac1{\kappa_n|a|}\gg1.
\end{align}
In a terminology of the renormalization group, this is a crossover from
the ultraviolet limit cycle governed by $a\to\infty$ to the infrared
limit cycle governed by $a\to0$.  If the Efimov effect does not take
place in the limit $a\to0$ which corresponds to
$\lambda_\mathrm{ir}\to\infty$, the Efimov spectrum of trimers in
Eq.~(\ref{eq:ultraviolet}) will terminate around $\kappa_n|a|\sim1$.

We demonstrate this general consideration by taking a particularly
interesting example in which two $A$ bosons are confined in 1D and
resonantly interact with one $B$ particle in 3D.  The zero-range
potential between $A$ bosons is given by
\begin{equation}
 V_A(z) = g_\mathrm{1D}\,\delta(z) \qquad \text{with} \qquad
  g_\mathrm{1D} \equiv -\frac{2\,\hbar^2}{m_Aa_\mathrm{1D}},
\end{equation}
where $a_\mathrm{1D}<0$ is the 1D scattering length.  At infinite
scattering length $a_\mathrm{1D}\to\infty$, the coupling vanishes
$g_\mathrm{1D}\to0$ and thus two $A$ bosons are noninteracting.
Therefore, the ultraviolet scaling factor
$\lambda_\mathrm{uv}=e^{\pi/s_\mathrm{uv}}$ is simply obtained from the
exponent $s_\mathrm{uv}=s_0$ for noninteracting $A$ bosons, which is
already plotted in the left panel of Fig.~\ref{fig:scaling_AAB} (dotted
curve) as a function of the mass ratio $m_A/m_B$.  On the other hand, at
zero scattering length $a_\mathrm{1D}\to-0$, the coupling diverges
$g_\mathrm{1D}\to+\infty$ and thus the potential $V_A(z)$ becomes a
hardcore repulsion.  Since the spectrum of bosons interacting via the
hardcore repulsion in 1D is equivalent to that of noninteracting
identical fermions~\cite{Girardeau:1960}, the infrared scaling factor
$\lambda_\mathrm{ir}=e^{\pi/s_\mathrm{ir}}$ is obtained from the
exponent $s_\mathrm{ir}=s_1$ for noninteracting $A$ ``fermions'', which
is also already plotted in the right panel of Fig.~\ref{fig:scaling_AAB}
(dotted curve).

\begin{figure*}
 \includegraphics[width=0.48\textwidth]{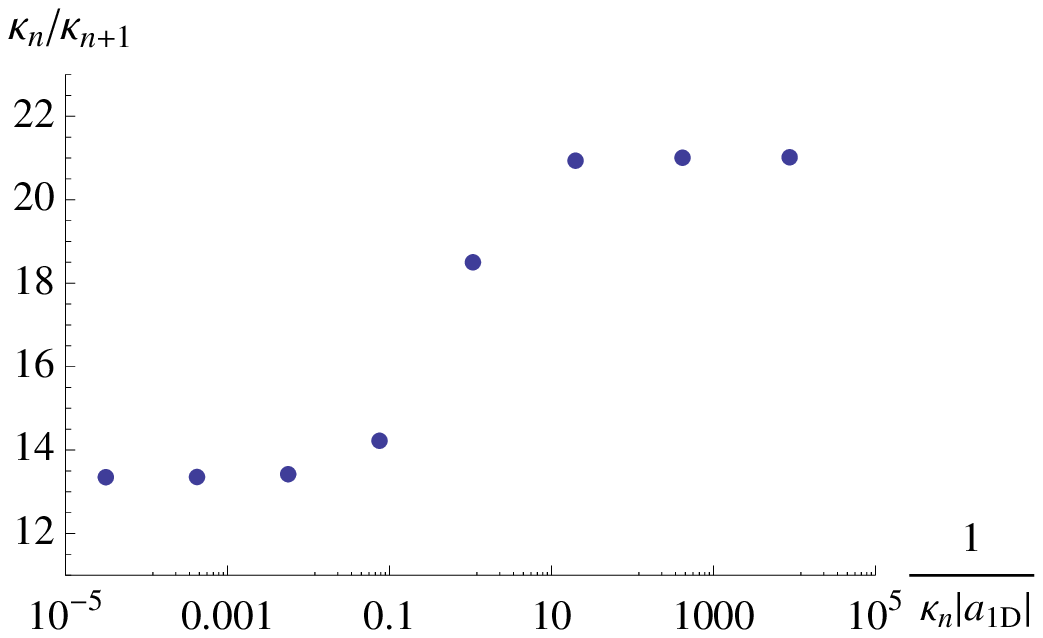}\hfill
 \includegraphics[width=0.48\textwidth]{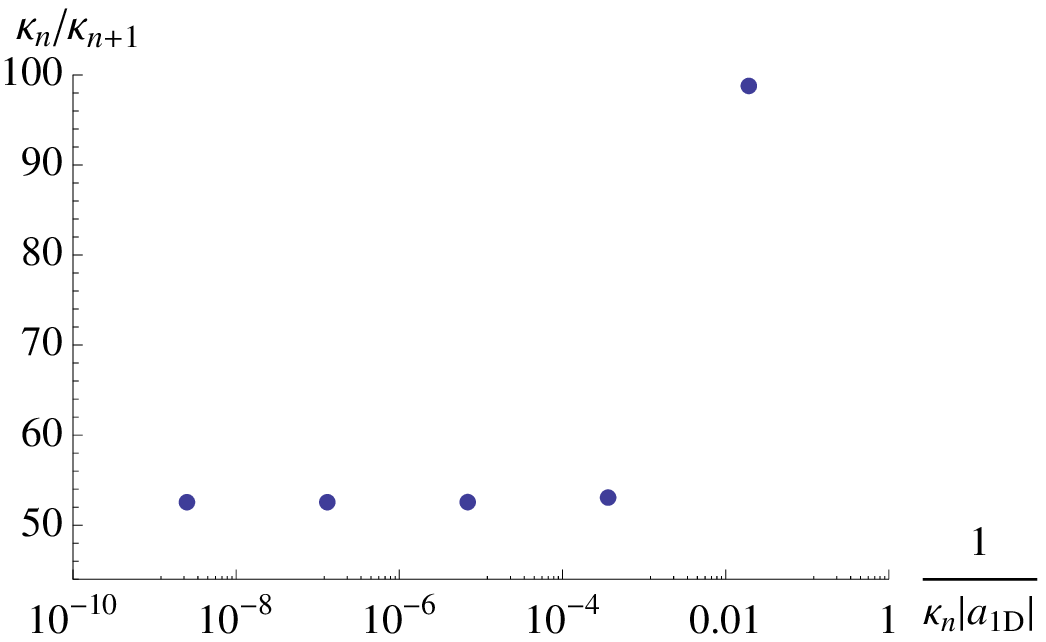}\smallskip
 \caption{Bose-Fermi crossover in ratios of two successive binding
 energies $\sqrt{E_n/E_{n+1}}=\kappa_n/\kappa_{n+1}$ of $AAB$ trimers in
 the even-parity channel as functions of $1/(\kappa_n|a_\mathrm{1D}|)$.
 The left and right panels are for mass ratios $m_A/m_B=41/6$ and
 $m_A/m_B=41/40$, respectively, and the 1D scattering length is chosen
 as $a_\mathrm{1D}=-\kappa_*^{-1}$, where $\kappa_*$ is the Efimov
 parameter in the noninteracting limit $a_\mathrm{1D}\to\infty$
 [see Eq.~(\ref{eq:boson})].  \label{fig:crossover}}
\end{figure*}

Figure~\ref{fig:crossover} shows the ratio of two successive binding
energies $\sqrt{E_n/E_{n+1}}=\kappa_n/\kappa_{n+1}$ in the even-parity
channel as a function of $1/(\kappa_n|a_\mathrm{1D}|)$ for two mass
ratios $m_A/m_B=41/6$ (left panel) and $m_A/m_B=41/40$ (right panel),
corresponding to $A={}^{41}$K, $B={}^6$Li and $A={}^{41}$K,
$B={}^{40}$K, respectively (see Appendix~\ref{app:crossover} for
details).  In the former case of $m_A/m_B=41/6$, the spectrum of $AAB$
trimers shows the crossover from the bosonic scaling behavior
\begin{align}
 \frac{\kappa_n}{\kappa_{n+1}} &\approx 13.3
 \qquad \text{for} \quad \frac1{\kappa_n|a_\mathrm{1D}|}\ll1
 \intertext{to the fermionic scaling behavior}
 \frac{\kappa_n}{\kappa_{n+1}} &\approx 21.0
 \qquad \text{for} \quad \frac1{\kappa_n|a_\mathrm{1D}|}\gg1.
\end{align}
On the other hand, in the latter case of $m_A/m_B=41/40$, the Efimov
spectrum of trimers
\begin{equation}
 \frac{\kappa_n}{\kappa_{n+1}} \approx 52.5
  \qquad \text{for} \quad  \frac1{\kappa_n|a_\mathrm{1D}|}\ll1
\end{equation}
terminates around $\kappa_n|a_\mathrm{1D}|\sim1$ because the mass ratio
is below the critical value of $m_A/m_B=2.06$ for $A$ ``fermions''.
This new phenomenon, {\em Bose-Fermi crossover\/} in the Efimov
spectrum, is unique in our 1D-3D mixture and will be in principle
observable in ultracold atom experiments.

\subsection[Interlayer Efimov trimers and stability in ultracold atoms]
  {Interlayer Efimov trimers and stability in ultracold atoms
  \cite{Nishida:2009nc} \label{sec:interlayer}}
Efimov trimers realized in ultracold atoms are unstable due to the
three-body recombination decaying into deeply bound dimers (see, for
example \cite{Ferlaino:2010}).  This would be a part of the reasons why
the spectroscopy of the Efimov trimers has not been performed in
ultracold atom experiments with the exception of recent
Refs.~\cite{Lompe:2010,Nakajima:2010}.  The idea of mixed dimensions is
useful to realize stable Efimov trimers in ultracold atoms.

In order for an $AAB$ trimer to decay into a deeply bound $AB$ dimer
with an $A$ atom, all three atoms have to come within a narrow range of
the interatomic potential $\sim r_0$.  This inelastic three-body
collision can be prevented by spatially separating two $A$ atoms by a
distance larger than $r_0$.  Therefore, we consider to confine two $A$
particles in two parallel 2D planes or 1D lines separated by a distance
$d$ and let them interact with one $B$ particle placed in 3D.  By making
$d\gg r_0$ and neglecting the interlayer or interwire tunneling, the
$AAB$ system becomes stable against the three-body recombination.
Furthermore, when the short-range interaction between $A$ and $B$
particles is resonant, two $A$ and one $B$ particles always form an
infinite tower of bound states even if two $A$ particles are spatially
separated regardless of their statistics and mass ratio.  This is
because spatially separated $A$ particles are distinguishable and also
because if the size of trimer
$\sim\kappa_n^{-1}\equiv\hbar/\sqrt{2\mu|E_n|}$ is much larger than $d$,
the separation of two planes or lines can be regarded as effectively
vanishing $d\to0$ and thus the problem reduces to the ordinary 2D-3D or
1D-3D mixture where the Efimov effect always appears in the $s$-wave or
even-parity channel ($\ell=0$) for any mass ratio as is shown in
Sect.~\ref{sec:confinement}.  Therefore, the spectrum of shallow $AAB$
trimers exhibits the approximate discrete scaling symmetry as
\begin{equation}\label{eq:interlayer}
 \frac{E_{n+1}}{E_n} \approx e^{-2\pi/s_0}
 \qquad \text{for} \qquad \kappa_n d\ll1,
\end{equation}
where the exponent $s_0$ is plotted in the left panel of
Fig.~\ref{fig:scaling_AAB} (dashed or dotted curve) as a function of the
mass ratio $m_A/m_B$.  On the other hand, since two $A$ particles are
spatially separated at least by the distance $d$, the size of trimer
$\sim\kappa_n^{-1}$ cannot be smaller than $d$ and thus the Efimov
spectrum of trimers in Eq.~(\ref{eq:interlayer}) has to be terminated by
a ground state trimer around $\kappa_0d\sim1$.

\begin{figure*}
 \includegraphics[width=0.48\textwidth]{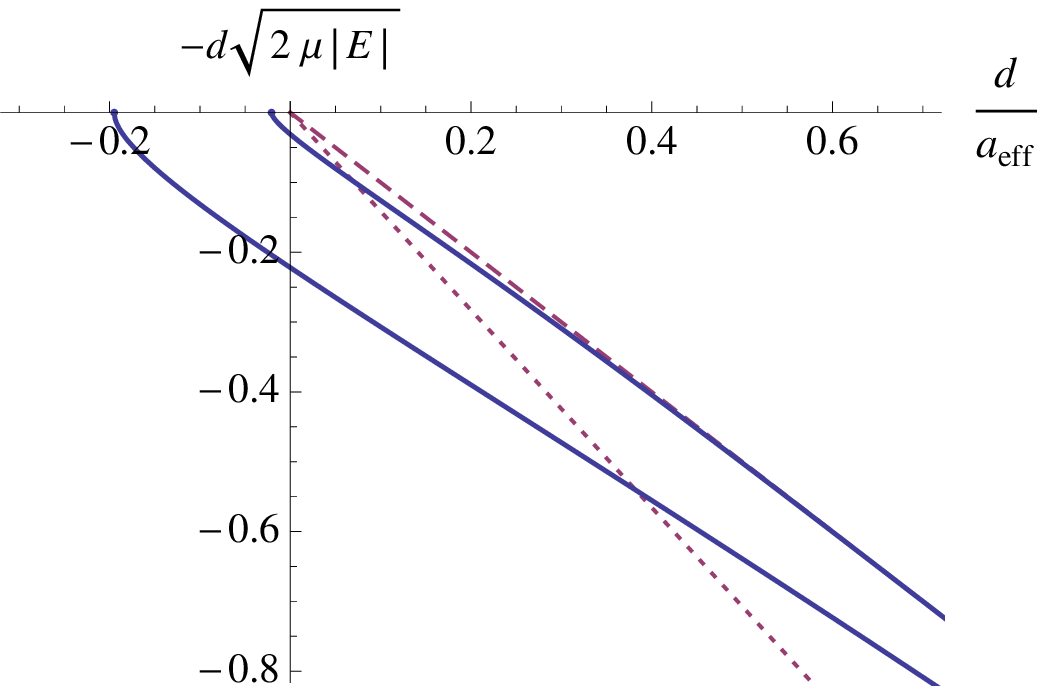}\hfill
 \includegraphics[width=0.48\textwidth]{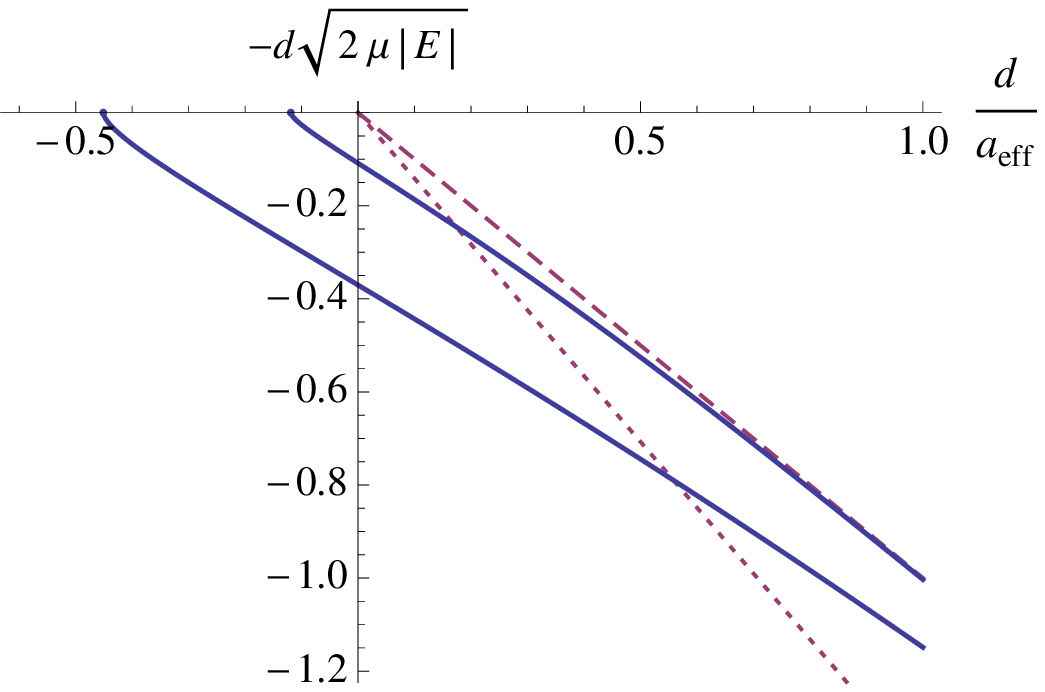}\smallskip
 \caption{Ground state binding energies $-d\sqrt{2\mu|E_0|/\hbar^2}$ of
 $AAB$ trimer in bilayer 2D-3D (left panel) and biwire 1D-3D (right
 panel) mixtures as functions of the inverse effective scattering length
 $d/a_\eff$.  The lower and upper solid curves are for mass ratios
 $m_A/m_B=40/6$ and $m_A/m_B=6/40$, respectively, and the dashed and
 dotted lines are atom-dimer and dimer-dimer thresholds;
 $E=-\hbar^2/(2\mu a_\eff^2)$ and $E=-\hbar^2/(\mu a_\eff^2)$.
 \label{fig:binding}}
\end{figure*}

Figure~\ref{fig:binding} shows the ground state binding energy
$-d\sqrt{2\mu|E_0|/\hbar^2}$ of $AAB$ trimer in bilayer 2D-3D (left
panel) and biwire 1D-3D (right panel) mixtures as a function of the
inverse effective scattering length $d/a_\eff$ for two mass ratios
$m_A/m_B=40/6$ and $m_A/m_B=6/40$, corresponding to $A={}^{40}$K,
$B={}^6$Li and $A={}^6$Li, $B={}^{40}$K,
respectively~\cite{Nishida:2009nc} (see Appendix~\ref{app:interlayer}
for details).\footnote{The analog of the scattering length in mixed
dimensions is referred to as the effective scattering length
$a_\eff$~\cite{Nishida:2008kr}.  In this article, we choose its
normalization so that the binding energy of two-body bound state for
$a_\eff>0$ is given by $E=-\hbar^2/(2\mu a_\eff^2)$.  See also
Refs.~\cite{Massignan:2006,Nishida:2010mw} in which the different
normalization was chosen so that $E=-\hbar^2/(2m_Ba_\eff^2)$.}  In
particular, the ground state binding energy in the unitarity limit
$a_\eff\to\infty$ is found to be
\begin{subequations}
 \begin{alignat}{2}
  E_0 &= -\frac{\hbar^2}{2\mu}\left(\frac{0.222}{d}\right)^2
  &\qquad &\text{for} \quad \frac{m_A}{m_B}=\frac{40}6 \\[8pt]
  E_0 &= -\frac{\hbar^2}{2\mu}\left(\frac{0.0310}{d}\right)^2
  &\qquad &\text{for} \quad \frac{m_A}{m_B}=\frac6{40}
 \end{alignat}
\end{subequations}
in bilayer 2D-3D mixture and
\begin{subequations}
 \begin{alignat}{2}
  E_0 &= -\frac{\hbar^2}{2\mu}\left(\frac{0.370}{d}\right)^2
  &\qquad &\text{for} \quad \frac{m_A}{m_B}=\frac{40}6 \\[8pt]
  E_0 &= -\frac{\hbar^2}{2\mu}\left(\frac{0.109}{d}\right)^2
  &\qquad &\text{for} \quad \frac{m_A}{m_B}=\frac6{40}
 \end{alignat}
\end{subequations}
in biwire 1D-3D mixture.\footnote{If we choose $d=390$\,nm for
$A={}^{40}$K and $B={}^6$Li, the ground state binding energy is given by
$E_0=-15.1$\,nK\,$\times\,k_B$ in bilayer 2D-3D mixture and
$E_0=-41.9$\,nK\,$\times\,k_B$ in biwire 1D-3D mixture.}  The ground state
trimer in the unitarity limit $a_\eff\to\infty$ is accompanied by an
infinite tower of excited trimer states (not shown in
Fig.~\ref{fig:binding}) exhibiting the discrete scaling symmetry as
\begin{subequations}
 \begin{alignat}{2}
  E_{n\gg1} &\approx -\frac{\hbar^2}{2\mu}
  \left(\frac{0.231}{d}\right)^2 \times \left(11.3\right)^{-2n}
  &\qquad &\text{for} \quad \frac{m_A}{m_B}=\frac{40}6 \\[8pt]
  E_{n\gg1} &\approx -\frac{\hbar^2}{2\mu}
  \left(\frac{0.0311}{d}\right)^2 \times \left(69.3\right)^{-2n}
  &\qquad &\text{for} \quad \frac{m_A}{m_B}=\frac6{40}
 \end{alignat}
\end{subequations}
in bilayer 2D-3D mixture and
\begin{subequations}
 \begin{alignat}{2}
  E_{n\gg1} &\approx -\frac{\hbar^2}{2\mu}
  \left(\frac{0.399}{d}\right)^2 \times \left(13.6\right)^{-2n}
  &\qquad &\text{for} \quad \frac{m_A}{m_B}=\frac{40}6 \\[8pt]
  E_{n\gg1} &\approx -\frac{\hbar^2}{2\mu}
  \left(\frac{0.109}{d}\right)^2 \times \left(76.6\right)^{-2n}
  &\qquad &\text{for} \quad \frac{m_A}{m_B}=\frac6{40}
 \end{alignat}
\end{subequations}
in biwire 1D-3D mixture.  Since the interlayer or interwire separation
$d$ is the only scale and plays a role of short-distance cutoff, the
spectrum of Efimov trimers that bridge between layers or wires is
completely determined by $d$ (and the effective scattering length
$a_\eff$ if finite).  Unlike ordinary Efimov trimers in a free space,
{\em these interlayer and interwire Efimov trimers are stable\/} in
ultracold atoms and thus will enable the spectroscopy of the Efimov
spectrum in future experiments.

As we have seen above, interlayer or interwire Efimov trimers in the
$s$-wave or even-parity channel ($\ell=0$) are always formed for any
mass ratio.  On the other hand, in the $p$-wave or odd-parity channel
($\ell=1$), they are formed only above a critical mass ratio given by
$m_A/m_B=6.35$ in bilayer 2D-3D mixture or $m_A/m_B=2.06$ in biwire
1D-3D mixture [see Eq.~(\ref{eq:critical})] and the scaling exponent
$s_1$ is plotted in the right panel of Fig.~\ref{fig:scaling_AAB}
(dashed or dotted curve) as a function of the mass ratio $m_A/m_B$.  In
the case of bilayer 2D-3D mixture, further infinite towers of interlayer
Efimov trimers can appear in higher partial-wave channels by increasing
$m_A/m_B$.

\subsection[Resonantly interacting anyons in two dimensions]
  {Resonantly interacting anyons in two dimensions \cite{Nishida:2007de}
  \label{sec:anyon}}
So far we have discussed the Efimov effect for resonantly interacting
bosons and fermions.  Here we discuss the possibility of the Efimov
effect for anyons in two dimensions.  As we have shown below
Eq.~(\ref{eq:constraint}) in Sect.~\ref{sec:absence}, the unitarity
limit becomes trivial both for bosons and fermions in two dimensions and
thus they cannot exhibit the Efimov effect.  On the other hand, the
unitarity limit is nontrivial for anyons and thus they may exhibit the
Efimov effect.

\begin{figure}
 \centering
 \includegraphics[width=0.48\textwidth]{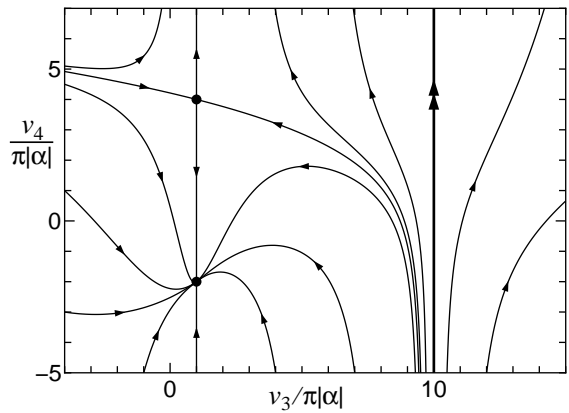}
 \caption{Renormalization group flow diagram in the plane of three-body
 and four-body couplings, $v_3$ and $v_4$, when the two-body interaction
 is tuned to the two-body resonance $a\to\infty$~\cite{Nishida:2007de}.
 If the three-body interaction is tuned to the ultraviolet fixed point
 at $v_3=10\pi|\alpha|$ corresponding to the three-body resonance, the
 four-body coupling $v_4$ exhibits the limit cycle (thick line with two
 arrows) indicating the Efimov effect in four anyons.
 \label{fig:RG_flow}}
\end{figure}

Ref.~\cite{Nishida:2007de} developed a perturbation theory to treat
anyons attracting via a resonant short-range interaction near the
fermionic limit.  It was found that unlike bosons in three dimensions,
three resonantly interacting anyons do not exhibit the Efimov effect.
However, if both two-body and three-body interactions are simultaneously
tuned to the two-body and three-body resonances, then the four-body
coupling exhibits the renormalization group limit cycle which indicates
{\em the Efimov effect in four anyons\/} (see Fig.~\ref{fig:RG_flow}).
Near the fermionic limit $|\alpha|\ll1$, the ratio of two successive
binding energies of four anyons shows the discrete scaling symmetry as
\begin{equation}
 \frac{E_{n+1}}{E_n}
  = \exp\!\left[-\frac{2\pi}{6|\alpha|+O(\alpha^2)}\right],
\end{equation}
where $\alpha$ is the statistics parameter defined so that $\alpha=0$
corresponds to fermions and $|\alpha|=1$ corresponds to bosons.  On the
other hand, anyons near the bosonic limit $|\alpha|\to1$ do not exhibit
the Efimov effect.  Therefore, the infinite tower of four-anyon bound
states found near the fermionic limit $|\alpha|\ll1$ has to disappear at
a certain critical value of $0<|\alpha|<1$.

\section{Summary \label{sec:summary}}
The Efimov physics has been studied for more than 40 years and
considered to exist only in three dimensions.  In this article, we
reviewed how the Efimov physics can be liberated from three dimensions.
We elucidated that scale-invariant interactions are necessary for the
Efimov effect and found that two-body and three-body interactions in
mixed dimensions and four-body interaction in one dimension become scale
invariant in the unitarity limit.  Then by adding another particle, all
such new systems indeed exhibit the Efimov effect.  Complete and unified
analyses of the Efimov effect in all possible cases will be presented in
a separate paper~\cite{Tan-Nishida}.  Here we focused on a number of
universal phenomena, such as confinement-induced Efimov effect,
Bose-Fermi crossover in Efimov spectrum, and formation of interlayer
Efimov trimers, that are unique in our new systems and observable in
ultracold atom experiments.  We also discussed the Efimov effect for
anyons in two dimensions.  Therefore, we can find the Efimov effect in
all spatial dimensions; three bosons with a resonant two-body
interaction in 3D~\cite{Efimov:1970}, four anyons with resonant two-body
and three-body interactions in 2D~\cite{Nishida:2007de}, five bosons
with a resonant four-body interaction in 1D~\cite{Nishida:2009pg}, and
various mixed-dimensional systems~\cite{Nishida:2008kr}.  An important
first step toward Efimov physics beyond three dimensions has been
recently taken by the experimental group at
Florence~\cite{Lamporesi:2010}.  They realized the 2D-3D mixed
dimensions by using the ultracold Bose-Bose mixture of ${}^{41}$K in 2D
and ${}^{87}$Rb in 3D and observed a series of two-body scattering
resonances where the Efimov effect is supposed to appear.  In view of
the fact that most studies on the Efimov effect have been devoted to
three dimensions for more than 40 years, we believe that this study
significantly broadens our horizons of universal Efimov physics.

\begin{acknowledgements}
 Y.\,N.\ was supported by MIT Pappalardo Fellowship in Physics and the
 U.\,S.\ Department of Energy under cooperative research agreement
 Contract Number DE-FG02-94ER40818.
\end{acknowledgements}

\appendix
\section{Details of Bose-Fermi crossover in 1D-3D mixture
 \label{app:crossover}}
Here we provide details of the Bose-Fermi crossover in Efimov spectrum
discussed in Sect.~\ref{sec:crossover}.  Two $A$ bosons in 1D
interacting with one $B$ particle in 3D are described by the
Schr\"odinger equation ($\hbar=1$):
\begin{equation}
 \left[-\frac1{2m_A}\left(\frac{\d^2}{\d z_{A1}^2}
 +\frac{\d^2}{\d z_{A2}^2}\right) - \frac1{2m_B}
 \left(\frac{\d^2}{\d z_B^2}+\frac{\d^2}{\d\x_B^2}\right)+V\right]
 \psi(z_{A1},z_{A2},z_B,\x_B) = E\,\psi(z_{A1},z_{A2},z_B,\x_B).
\end{equation}
Here $\x=(x,y)$ are two-dimensional coordinates and the zero-range
potential $V$ is given by
\begin{equation}
 \begin{split}
  V \psi(z_{A1},z_{A2},z_B,\x_B)
  &= \delta(z_B-z_{A1})\,\delta\Bigl(\sqrt{\tfrac{m_B}{\mu}}\x_B\Bigr)\,f(z_{A1}-z_{A2})
  + \delta(z_B-z_{A2})\,\delta\Bigl(\sqrt{\tfrac{m_B}{\mu}}\x_B\Bigr)\,f(z_{A2}-z_{A1}) \\
  &\quad + g_\mathrm{1D}\,\delta(z_{A1}-z_{A2})\,\psi(z_{A1},z_{A1},z_B,\x_B).
 \end{split}
\end{equation}
The unknown function $f$ is defined as
\begin{equation}\label{eq:unknown}
 f(z_{A1}-z_{A2}) \equiv \frac{2\pi a_\eff}{\mu}
  \lim_{z_B\to z_{A1},\x_B\to\0}\frac{\d}{\d R}
  \left[R\,\psi(z_{A1},z_{A2},z_B,\x_B)\right],
\end{equation}
where $\mu\equiv m_Am_B/(m_A+m_B)$ is the reduced mass, $a_\eff$ is the
effective scattering length, and
$R\equiv\sqrt{(z_B-z_{A1})^2+\frac{m_B}{\mu}\x_B^2}$ is the ``distance''
between $A$ and $B$ particles~\cite{Nishida:2008kr}.  Since the system
is translationally invariant in the $z$ direction, it is convenient to
introduce new coordinates
\begin{equation}
 z_{12} \equiv z_{A1}-z_{A2}, \qquad
  z_{AB} \equiv z_B-\frac{z_{A1}+z_{A2}}2,
  \qquad \text{and} \qquad
  Z \equiv \frac{m_Az_{A1}+m_Az_{A2}+m_Bz_B}{2m_A+m_B}
\end{equation}
and separate the center-of-mass coordinate $Z$.

For bound states $E=-|E|<0$, the Schr\"odinger equation is formally
solved by
\begin{equation}
 \psi(z_{12},z_{AB},\x_B) = -\int dz'_{12}dz'_{AB}d\x'_B\,
  \<z_{12},z_{AB},\x_B|\frac1{H_0+|E|}|z'_{12},z'_{AB},\x'_B\>
  V\psi(z'_{12},z'_{AB},\x'_B).
\end{equation}
By imposing the short-range boundary condition from
Eq.~(\ref{eq:unknown})
\begin{equation}
 \lim_{z_{AB}\to\frac{z_{12}}2,\x_B\to\0}\psi(z_{12},z_{AB},\x_B)
  = \frac{\mu}{2\pi}\left[\frac1{a_\eff}-\frac1{R}\right]f(z_{12}) + O(R)
\end{equation}
and by taking the limit $z_{12}\to0$, we obtain a set of two integral
equations obeyed by $f(z_{12})$ and
$g(z_{AB},\x_B)\equiv g_\mathrm{1D}\psi(0,z_{AB},\x_B)$.  In the
unitarity limit $a_\eff\to\infty$, these integral equations in the
momentum space are expressed by
\begin{equation}
 \begin{split}
  & \frac{\mu}{m_B}\int\!\frac{dq_zd\q}{(2\pi)^3}
  \left[\frac1{\frac{\left(2p_z-q_z\right)^2}{4m_A}
  +\frac{2m_A+m_B}{4m_Am_B}q_z^2+\frac{\q^2}{2m_B}+|E|}
  - \frac1{\frac{q_z^2}{2\mu}+\frac{\q^2}{2m_B}}\right]f(p_z) \\
  &= -\frac{\mu}{m_B}\int\!\frac{dq_zd\q}{(2\pi)^3}
  \frac{f(q_z-p_z)}{\frac{\left(2p_z-q_z\right)^2}{4m_A}
  +\frac{2m_A+m_B}{4m_Am_B}q_z^2+\frac{\q^2}{2m_B}+|E|}
  - \int\!\frac{dq_zd\q}{(2\pi)^3}\frac{g(q_z,\q)}
  {\frac{\left(2p_z-q_z\right)^2}{4m_A}
  +\frac{2m_A+m_B}{4m_Am_B}q_z^2+\frac{\q^2}{2m_B}+|E|}
 \end{split}
\end{equation}
and
\begin{equation}
 \begin{split}
  \left[\frac1{g_\mathrm{1D}} + \int\!\frac{dq_z}{2\pi}\frac1{\frac{q_z^2}{m_A}
  +\frac{2m_A+m_B}{4m_Am_B}p_z^2+\frac{\p^2}{2m_B}+|E|}\right]g(p_z,\p)
  &= -\frac{2\mu}{m_B}\int\!\frac{dq_z}{2\pi}\frac{f(q_z+\tfrac{p_z}2)}
  {\frac{q_z^2}{m_A}+\frac{2m_A+m_B}{4m_Am_B}p_z^2+\frac{\p^2}{2m_B}+|E|}.
 \end{split}
\end{equation}
By substituting the solution $g(p_z,\p)$ of the second equation into the
first integral equation, we arrive at a closed integral equation obeyed
by $f(p_z)$:
\begin{equation}
 \begin{split}
  \sqrt{\frac{2u+1}{(u+1)^2}p_z^2+\varepsilon}\,f(p_z)
  &= 4\pi\int\limits^\Lambda\frac{dq_zd\q}{(2\pi)^3}
  \frac{f(q_z)}{p_z^2+q_z^2+\frac{2u}{u+1}p_zq_z+\q^2+\varepsilon} \\
  &\quad -16\pi\sqrt{\frac2{u+1}}\int\!\frac{dk_zd\k}{(2\pi)^3}
  \frac1{p_z^2+\sqrt{\frac2{u+1}}p_zk_z+k_z^2+\k^2+\frac{u+1}2\varepsilon} \\
  &\qquad \times \frac1{|a_\mathrm{1D}|
  +\frac1{\sqrt{\frac{2u+1}{2u+2}k_z^2+\k^2+\frac{u+1}2\varepsilon}}}
  \int\limits^\Lambda\frac{dq_z}{2\pi}\frac{f(q_z)}
  {q_z^2+\sqrt{\frac2{u+1}}q_zk_z+k_z^2+\k^2+\frac{u+1}2\varepsilon},
 \end{split}
\end{equation}
where we defined $u\equiv m_A/m_B$ and $\varepsilon\equiv2\mu|E|$ and
rescaled integration variables to simplify the expression.  This
integral equation in the even-parity channel $f(p_z)=f(-p_z)$ has to be
solved numerically to find the spectrum of $AAB$ trimers.

In the noninteracting limit $a_\mathrm{1D}\to\infty$, we find the Efimov
spectrum
\begin{equation}\label{eq:boson}
 E_n = -\frac{\hbar^2\kappa_*^2}{2\mu}\times e^{-2\pi n/s_\mathrm{uv}},
\end{equation}
where the exponent $s_\mathrm{uv}$ is an imaginary part of the scaling
exponent $\gamma_\ell$ solving the following equation in the even-parity
channel ($\ell=0$):
\begin{equation}
 \frac{\sqrt{2u+1}}{u+1}
  = -\frac{\cos[(\gamma_\ell+1)\arccos(\frac{u}{u+1})]
  + (-1)^\ell\cos[(\gamma_\ell+1)\arccos(\frac{-u}{u+1})]}
  {(\gamma_\ell+1)\sin[(\gamma_\ell+1)\pi]}.
\end{equation}
On the other hand, in the hardcore limit $a_\mathrm{1D}\to-0$, we again
find the Efimov spectrum
\begin{equation}\label{eq:fermion}
 E_n = -\frac{\hbar^2\kappa_*'^2}{2\mu}\times e^{-2\pi n/s_\mathrm{ir}},
\end{equation}
where the exponent $s_\mathrm{ir}$ is an imaginary part of $\gamma_\ell$
solving the following equation in the even-parity channel ($\ell=0$):
\begin{equation}\label{eq:hardcore}
 \frac{\sqrt{2u+1}}{u+1}
  = \frac{\cos[(\gamma_\ell+1)\arccos(\frac{u}{u+1})]
  - (-1)^\ell\cos[(\gamma_\ell+1)\arccos(\frac{-u}{u+1})]}
  {(\gamma_\ell+1)\sin[(\gamma_\ell+1)\pi]}.
\end{equation}
We note that the scaling exponent $\gamma_\ell$ of noninteracting $A$
fermions satisfies the same equation (\ref{eq:hardcore}) by exchanging
the roles of even and odd parities $\ell=0\leftrightarrow1$.  Therefore,
the hardcore $A$ bosons have the same energy spectrum as the
noninteracting $A$ ``fermions''.  When $0<|a_\mathrm{1D}|<\infty$, the
spectrum of $AAB$ trimers shows the crossover from the bosonic scaling
behavior (\ref{eq:boson}) for
$\sqrt{2\mu|E_n|/\hbar^2}|a_\mathrm{1D}|\gg1$ to the fermionic scaling
behavior (\ref{eq:fermion}) for
$\sqrt{2\mu|E_n|/\hbar^2}|a_\mathrm{1D}|\ll1$.  This Bose-Fermi
crossover can be seen clearly in Fig.~\ref{fig:crossover}, where the
ratio of two successive binding energies
$\sqrt{E_n/E_{n+1}}=\kappa_n/\kappa_{n+1}$ is plotted as a function of
$1/(\kappa_n|a_\mathrm{1D}|)$ for two mass ratios $m_A/m_B=41/6$ (left
panel) and $m_A/m_B=41/40$ (right panel) by choosing the 1D scattering
length as $a_\mathrm{1D}=-\kappa_*^{-1}$.

\section{Details of interlayer and interwire Efimov trimers
 \label{app:interlayer}}
Here we provide details of the formation of interlayer and interwire
Efimov trimers discussed in Sect.~\ref{sec:interlayer}.

\subsection{Bilayer 2D-3D mixture}
Two $A$ particles in two parallel 2D planes placed at $z=z_{A1}$ and
$z_{A2}$ interacting with one $B$ particle in 3D are described by the
Schr\"odinger equation ($\hbar=1$):
\begin{equation}
 \left[-\frac1{2m_A}\left(\frac{\d^2}{\d\x_{A1}^2}
 +\frac{\d^2}{\d\x_{A2}^2}\right) - \frac1{2m_B}
 \left(\frac{\d^2}{\d\x_B^2}+\frac{\d^2}{\d z_B^2}\right)+V\right]
 \psi(\x_{A1},\x_{A2},\x_B,z_B) = E\,\psi(\x_{A1},\x_{A2},\x_B,z_B).
\end{equation}
Here $\x=(x,y)$ are two-dimensional coordinates and the zero-range
potential $V$ is given by
\begin{equation}
 \begin{split}
  V \psi(\x_{A1},\x_{A2},\x_B,z_B) &= \delta(\x_B-\x_{A1})\,
  \delta\Bigl(\sqrt{\tfrac{m_B}{\mu}}(z_B-z_{A1})\Bigr)\,f_1(\x_{A1}-\x_{A2}) \\
  &\quad + \delta(\x_B-\x_{A2})\,
  \delta\Bigl(\sqrt{\tfrac{m_B}{\mu}}(z_B-z_{A2})\Bigr)\,f_2(\x_{A2}-\x_{A1}).
 \end{split}
\end{equation}
The unknown functions $f_1$ and $f_2$ are defined as
\begin{equation}\label{eq:unknown_1}
 f_1(\x_{A1}-\x_{A2}) \equiv \frac{2\pi a_\eff}{\mu}
  \lim_{\x_B\to\x_{A1},z_B\to z_{A1}}\frac{\d}{\d R_1}
  \left[R_1\psi(\x_{A1},\x_{A2},\x_B,z_B)\right]\phantom{,}
\end{equation}
and
\begin{equation}\label{eq:unknown_2}
 f_2(\x_{A2}-\x_{A1}) \equiv \frac{2\pi a_\eff}{\mu}
  \lim_{\x_B\to\x_{A2},z_B\to z_{A2}}\frac{\d}{\d R_2}
  \left[R_2\psi(\x_{A1},\x_{A2},\x_B,z_B)\right],
\end{equation}
where $\mu\equiv m_Am_B/(m_A+m_B)$ is the reduced mass, $a_\eff$ is the
effective scattering length, and
$R_i\equiv\sqrt{(\x_B-\x_{Ai})^2+\frac{m_B}{\mu}(z_B-z_{Ai})^2}$
\ ($i=1,2$) is the ``distance'' between $A$ and $B$
particles~\cite{Nishida:2008kr}.  Since the system is translationally
invariant in the $x$ and $y$ directions, it is convenient to introduce
new coordinates
\begin{equation}
 \x_{12} \equiv \x_{A1}-\x_{A2}, \qquad
  \x_{AB} \equiv \x_B-\frac{\x_{A1}+\x_{A2}}2,
  \qquad \text{and} \qquad
  \bm{X} \equiv \frac{m_A\x_{A1}+m_A\x_{A2}+m_B\x_B}{2m_A+m_B}
\end{equation}
and separate the center-of-mass coordinate $\bm{X}$.

For bound states $E=-|E|<0$, the Schr\"odinger equation is formally
solved by
\begin{equation}
 \psi(\x_{12},\x_{AB},z_B) = -\int d\x'_{12}d\x'_{AB}dz'_B\,
  \<\x_{12},\x_{AB},z_B|\frac1{H_0+|E|}|\x'_{12},\x'_{AB},z'_B\>
  V\psi(\x'_{12},\x'_{AB},z'_B).
\end{equation}
By imposing the short-range boundary conditions from
Eq.~(\ref{eq:unknown_1})
\begin{equation}
 \lim_{\x_{AB}\to\frac{\x_{12}}2,z_B\to z_{A1}}\psi(\x_{12},\x_{AB},z_B)
  = \frac{\mu}{2\pi}\left[\frac1{a_\eff}-\frac1{R_1}\right]f_1(\x_{12}) + O(R_1)
\end{equation}
and from Eq.~(\ref{eq:unknown_2})
\begin{equation}
 \lim_{\x_{AB}\to-\frac{\x_{12}}2,z_B\to z_{A2}}\psi(\x_{12},\x_{AB},z_B)
  = \frac{\mu}{2\pi}\left[\frac1{a_\eff}-\frac1{R_2}\right]f_2(-\x_{12}) + O(R_2),
\end{equation}
we obtain a set of two integral equations obeyed by $f_1(\x_{12})$ and
$f_2(\x_{12})$.  These integral equations in the momentum space are
expressed by
\begin{equation}
 \begin{split}
  & \left[\frac{\mu}{2\pi a_\eff}
  + \sqrt{\frac{\mu}{m_B}}\int\!\frac{d\q dq_z}{(2\pi)^3}
  \left(\frac1{\frac{\left(2\p-\q\right)^2}{4m_A}
  +\frac{2m_A+m_B}{4m_Am_B}\q^2+\frac{q_z^2}{2m_B}+|E|}
  - \frac1{\frac{\q^2}{2\mu}
  +\frac{q_z^2}{2m_B}}\right)\right]f_1(\p) \\
  &= -\sqrt{\frac{\mu}{m_B}}\int\!\frac{d\q dq_z}{(2\pi)^3}
  \frac{e^{iq_z(z_{A1}-z_{A2})}}{\frac{\left(2\p-\q\right)^2}{4m_A}
  +\frac{2m_A+m_B}{4m_Am_B}\q^2+\frac{q_z^2}{2m_B}+|E|}f_2(\q-\p)
 \end{split}
\end{equation}
and $1\leftrightarrow2$.  By defining $f_\pm=f_1\pm f_2$, we arrive at a
closed integral equation obeyed by $f_\pm(\p)$:
\begin{equation}
 \left[\sqrt{\frac{2u+1}{(u+1)^2}\p^2+\varepsilon}
  - \frac1{a_\eff}\right]f_\pm(\p)
 = \pm\int\!\frac{d\q}{2\pi}
 \frac{e^{-d\sqrt{\frac{u+1}{u}}\sqrt{\p^2+\q^2+\frac{2u}{u+1}\p\cdot\q+\varepsilon}}}
 {\sqrt{\p^2+\q^2+\frac{2u}{u+1}\p\cdot\q+\varepsilon}}f_\pm(\q),
\end{equation}
where we defined $u\equiv m_A/m_B$, $\varepsilon\equiv2\mu|E|$, and the
interlayer separation $d\equiv|z_{A1}-z_{A2}|$.  Finally, the
partial-wave projection
$f_\pm^{(\ell)}(p)\equiv\int_0^\pi\!\frac{d\varphi}\pi\,\cos(\ell\varphi)\,f_\pm(\p)$
leads to
\begin{equation}
 \left[\sqrt{\frac{2u+1}{(u+1)^2}p^2+\varepsilon}
  - \frac1{a_\eff}\right]f_\pm^{(\ell)}(p)
 = \pm\int_0^\infty\!dq\,q\int_0^\pi\!\frac{d\varphi}\pi\,\cos(\ell\varphi)\,
 \frac{e^{-d\sqrt{\frac{u+1}{u}}\sqrt{p^2+q^2+\frac{2u\cos\varphi}{u+1}pq+\varepsilon}}}
 {\sqrt{p^2+q^2+\frac{2u\cos\varphi}{u+1}pq+\varepsilon}}f_\pm^{(\ell)}(q).
\end{equation}
The same integral equation has been derived in
Ref.~\cite{Nishida:2009nc} by using a field theoretical method.  This
integral equation for $f_+^{(\ell)}$ in the $s$-wave channel ($\ell=0$)
has been solved numerically to find the spectrum of interlayer $AAB$
trimers which is plotted as a function of $d/a_\eff$ in the left panel
of Fig.~\ref{fig:binding} for two mass ratios $m_A/m_B=40/6$ and
$m_A/m_B=6/40$.

\subsection{Biwire 1D-3D mixture}
The case for two $A$ particles in two parallel 1D lines placed at
$\x=\x_{A1}$ and $\x_{A2}$ interacting with one $B$ particle in 3D can
be studied in the same way by exchanging the roles of above
$\x\leftrightarrow z$.  The resulting integral equation obeyed by
$f_\pm(p_z)\equiv f_1(p_z)\pm f_2(p_z)$ is
\begin{equation}
 \left[\sqrt{\frac{2u+1}{(u+1)^2}p_z^2+\varepsilon}
  - \frac1{a_\eff}\right]f_\pm(p_z)
 = \pm\int_{-\infty}^\infty\!\frac{dq_z}\pi
 K_0\!\left(d\sqrt{\frac{u+1}{u}}\sqrt{p_z^2+q_z^2
       +\frac{2u}{u+1}p_zq_z+\varepsilon}\right)f_\pm(q_z),
\end{equation}
where we defined the interwire separation $d\equiv|\x_{A1}-\x_{A2}|$.
This integral equation for $f_+$ in the even-parity channel
$f(p_z)=f(-p_z)$ has been solved numerically to find the spectrum of
interwire $AAB$ trimers which is plotted as a function of $d/a_\eff$ in
the right panel of Fig.~\ref{fig:binding} for two mass ratios
$m_A/m_B=40/6$ and $m_A/m_B=6/40$.

\end{document}